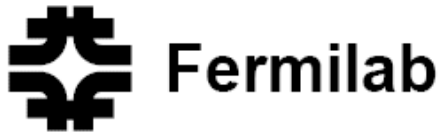

# SHIELDING EXPERIMENTS UNDER JASMIN COLLABORATION AT FERMILAB (IV) – MEASUREMENT AND ANALYSES OF HIGH-ENERGY NEUTRON SPECTRA IN THE ANTI-PROTON TARGET STATION[*,†]

N. MATSUDA[#,1] Y. KASUGAI[1], H. MATSUMURA[2], H. YASHIMA[3], H. IWASE[2], N. KINOSHITA[2,a], N. MOKHOV[4], A. LEVELING[4], D. BOEHNLEIN[4], K. VAZIRI[4], G. LAUTEN[4], W. SCHMITT[4], K. OISHI[5], T. NAKAMURA[6], H. HIRAYAMA[2], K. ISHIBASHI[7], K. NIITA[8], Y. SAKAMOTO[1], and H. NAKASHIMA[1]

[1]Japan Atomic Energy Agency, 2-4 Shirakata Shirane, Tokai-mura, Ibaraki 319-1195, Japan
[2]High Energy Accelerator Organization, 1-1 Oho, Tsukuba, Ibaraki 305-0801, Japan
[3]Research Reactor Institute, Kyoto University, Kumatori-cho, Sennan-gun, Osaka 590-0494, Japan
[4]Fermi National Accelerator Laboratory, Batavia IL 60510, USA
[5]Shimizu Corporation, 4-17, Echujima 3-chome, Koto-ku, Tokyo 135-8530, Japan
[6]Tohoku University, Aoba, Aramaki, Aoba-ku, Sendai, Miyagi 980-8578, Japan
[7]Kyushu University, 744, Motooka, Nishi-ku, Fukuoka 819-0935, Japan
[8]Research Organization for information Science & Technology, 2-4 Shirakata Shirane, Tokai-mura, Ibaraki 319-1106, Japan

## Abstract

Neutron spectra in high-energy region between 1 and 100 MeV in the shield configuration of the anti-proton target station and a 120-GeV proton beam at Fermi National Accelerator Laboratory (Fermilab) were determined using the reaction rate data obtained with the multi-foil activation method. Two kinds of methods were employed for the determination of neutron spectra: one is the fitting method which is newly developed in this work, another is the unfolding method with SAND-II code. The calculations were performed using the PHITS. From the comparison between the calculated and experimental neutron spectra, it concluded that the PHITS can be used for shielding design of high-energy proton accelerators.

[*]Work supported by grant aid of the Ministry of Education (KAKENHI 19360432 and 20354764) in Japan and by Fermi Research Alliance, LLC under contract No. DE-AC02-07CH11359 with the U.S. Department of Energy.
[†]Published in *Journal of the Korean Physical Society*.
[#]Corresponding author. E-mail: matsuda.norihiro@jaea.go.jp
[a]Present address : University of Tsukuba, Tennodai 1-1-1, Tsukuba, Ibaraki 305-8577, Japan

## 1. INTRODUCTION

The Japanese and American Study of Muon Interaction and Neutron Detection (JASMIN) collaboration [1] has been organized to study radiation safety issues on ultra high-energy accelerators. The experiments have been carried out at Fermi National Accelerator Laboratory (Fermilab). The main purpose of this study is to perform systematic acquisition of shielding data for high-energy accelerators with incident proton-beam energies above 100 GeV. One of the goals of the study is to measure the associated secondary neutron energy spectra. This work presents the results of neutron-energy spectra measurements in the energy region from 1 to 100 MeV at various locations in the target area shielding. The measured neutron spectra were compared with calculation results by PHITS code [2] for benchmarking.

## 2. Experiment

The experiments were carried out at the anti-proton target station (Pbar) of Fermilab. The Pbar facility is used for anti-proton production by bombarding an inconel target with 120-GeV protons. The target is surrounded by thick iron and concrete shielding, as shown in Fig. 1.

In order to determine high-energy neutron spectra inside the shielding, the multi-foil activation technique was utilized. Activation detectors of aluminium, niobium, indium and bismuth were installed in the shield assembly as shown in Fig. 1. After the irradiations these foils were analyzed by measuring the reaction rates of $^{27}$Al(n, α)$^{24}$Na, $^{209}$Bi(n, xn)$^{203-206}$Bi ($x$ = 4~7), $^{93}$Nb(n, 2n)$^{92m}$Nb and $^{115}$In(n, n')$^{115m}$In. These reactions are sensitive to neutron energies between 0.1 and 100 MeV. The reaction rates were determined by gamma-ray spectrometry using high-purity germanium detectors. The details of the experimental procedure are given in reference [3].

The reaction rate data include much information such as neutron attenuation by the shield material and flux distributions, which are described in the first paper in this series [4]. Moreover, in this work, we determined neutron energy spectra by using the differences of the neutron responses among various threshold reactions. The determination methods are described in the next section.

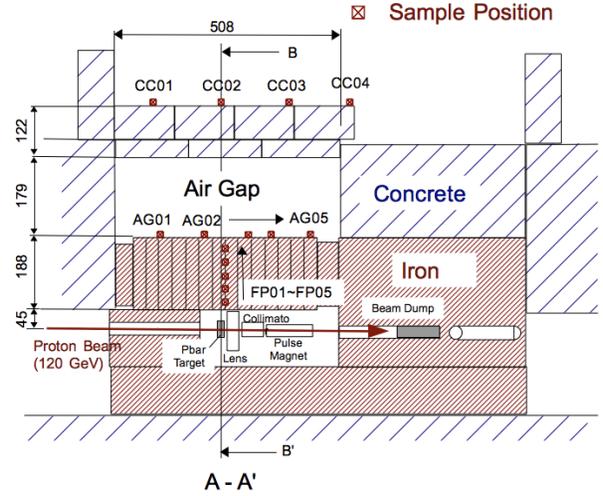

**Fig. 1.** Cross sectional view of the anti-proton target station at Fermilab. Dimensions are shown in units of cm. The samples for activation were set at the positions marked as squares with crosses. Foil positions were labelled as FP01-FP05 (from bottom to top), AG01-AG05 and CC01-CC04 (in the proton beam ddirection). FP means "Filler Plate" used for mounting the samples in the steel shied. AG and CC is abbreviations of "Air Gap" and "Concrete Cap", respectively.

## 3. DETERMINATION OF NEUTRON SPECTRA

In this section, based on the reaction rate data, neutron spectrum between 1 to 100 MeV at each location was determined using two method: fitting method and the unfolding method. The fitting method has been recently developed for this work. The details of each method are described in the following sub-sections. In the sub-section 3.3, the neutron spectra determined with both methods are compared with each other, and their validity is discussed.

### 3.1 Fitting Method

We assumed that neutron energy spectra between 1 and 100 MeV could be expressed with a as a sum of the two exponential functions

$$\phi_{fit}(E) = a_1 \exp(-a_2 E) + a_3 \exp(-a_4 E), \quad (1)$$

where $\phi$ is a neutron flux in unit of MeV$^{-1}$, $E$ is the neutron energy and $a_1$~ $a_4$ are fitting parameters, which are determined with the non-linear least-squares fit to the extant data. An example of the fitting function is shown with each exponential component in Fig. 2. The spectrum is constructed by smoothly connected both simplified 1/$E$ spectrum and the evaporation peak within the range. The fitting parameters were adjusted to minimize χ$^2$-values defined as

$$\chi^2 = \sum_i \left( \frac{R^i_{fit} - R^i_{exp}}{\varepsilon^i} \right)^2, \qquad (2)$$

where $R^i_{fit}$ and $R^i_{exp}$ are fitting and experimental reaction rates of the $i$th-reaction, and $\varepsilon^i$ is the experimental error of the $i$th-reaction. The values of $R^i_{fit}$ were calculated using

$$R^i_{fit} = \sum_j \sigma^i(E_j) \phi_{fit}(E_j) \Delta E_j. \qquad (3)$$

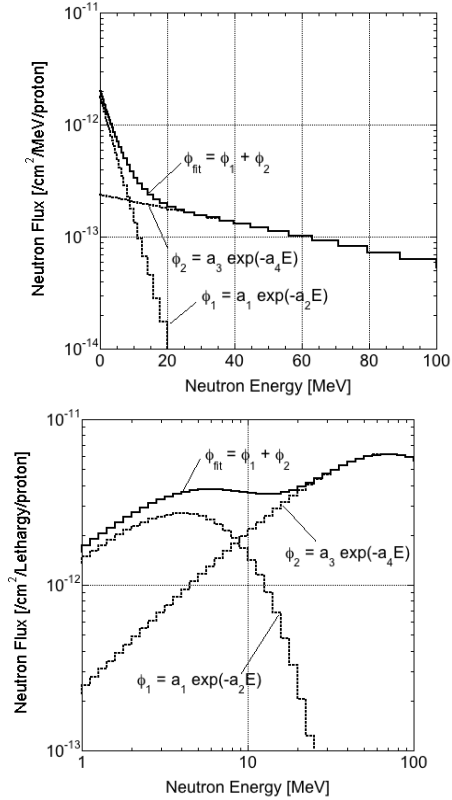

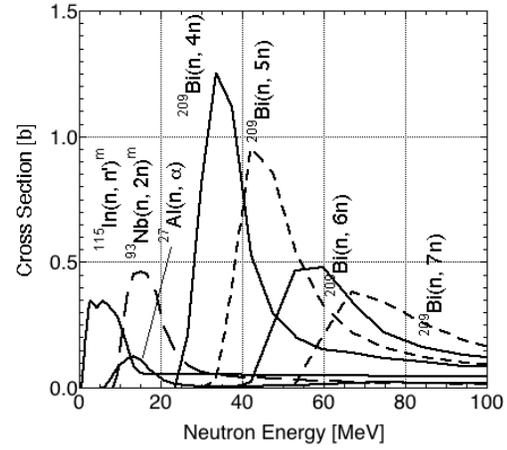

**Fig. 2.** Shape of the neutron spectrum assumed for fitting method. The neutron spectra in the upper and lower figures are drawn in the unit of MeV$^{-1}$ and lethargy$^{-1}$, respectively. The two exponential components are also shown in each figure.

In formula (3), $\sigma^i(E_j)$ is the cross section of the $i$th-reaction at a neutron energy of $E_j$, and $\Delta E_j$ is the width of neutron-energy bin at $E_j$.

The cross section curves used for determination of neutron spectra are shown in Fig. 3. The cross sections [5] were originally evaluated in 1999 on the basis of existing experimental data and evaluations.

**Fig. 3.** Cross-sections [5] used for determination of neutron spectra. These cross-sections were originally evaluated based on the existing experimental data and theoretical calculations.

### 3.2 Unfolding Method

Unfolding was carried out using SAND-II code [6]. As initial guess spectra, the results obtained with the fitting method were adopted. The cross-sections shown in Fig. 3 were also used for the unfolding.

### 3.3 Results

Neutron spectrum at each sample location in Fig. 1 was determined with the two methods. For example, the spectra at CC01~CC04 on top of the concrete shield are shown in Fig. 4. In the figure, both of the fitting and unfolding spectra are shown for comparisons. In Fig. 5, the ratios of reaction rates calculated from fitting and unfolding neutron spectra to the experimental values are shown for all reactions used to determine the neutron spectra. Though the detailed structure of a neutron spectrum could not be expressed by the fitting function, the C/E-values of for the fitting method were satisfactory, taking the experimental errors into account. This indicates that the fitting function (1) which we adopted was valid for expressing a neutron spectrum in this energy region. The C/E-values for the unfolding spectra were closer to 1 than those for the fitting method. It should be noted that the peaks and valleys between 30 and 70 MeV in the unfolding spectra are probably due to the inconsistencies in the cross-section data for $^{209}$Bi(n, $x$n)$^{203-207}$Bi reactions. We presume that these cross section data do not have enough accuracy for requirement of neutron dosimetry, and further experimental data will be needed for a more precise evaluation. Therefore, in this study, we choose the results of the fitting method for further discussion.

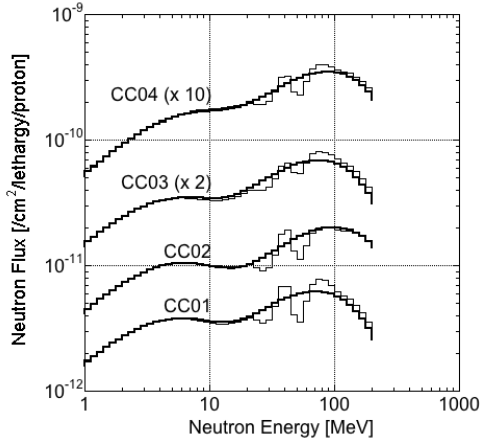

**Fig. 4.** Neutron spectra determined with the fititng and undolding methods at the samples locations of CC01-CC04 on the top of the concrete shield. The thick black lines and fine lines show the neutron spectra determined using the fitting and unfolding methods, respectively.

the devices in the beam tunnel such as a pulse magnet and the shield configuration were simplified. The calculated neutron spectra at 90° to the proton beam direction are shown with the experimental ones determined with the fitting method for in Fig. 6. The experimental neutron spectrum on the top of the steel shield at 90°, shown as AG($\theta$=90°) in Fig. 6, was determined by interpolating the reaction rate data at AG01 and AG02 since there were no measurements at $\theta$=90° in the Air Gap. The shapes of the calculated neutron spectra at AG($\theta$=90°) and CC02 show good agreement with the experimental ones. On the whole, the calculations tend to overestimate the neutron fluxes, and the differences increase with increase in shielding thickness. This means that the calculation underestimates neutron attenuation lengths of steel and concrete. Since the calculated neutron fluxes are reasonably more conservative from the view point of radiation shield engineering, it can be conclude that the PHITS is suitable for shielding design of high-energy proton accelerators.

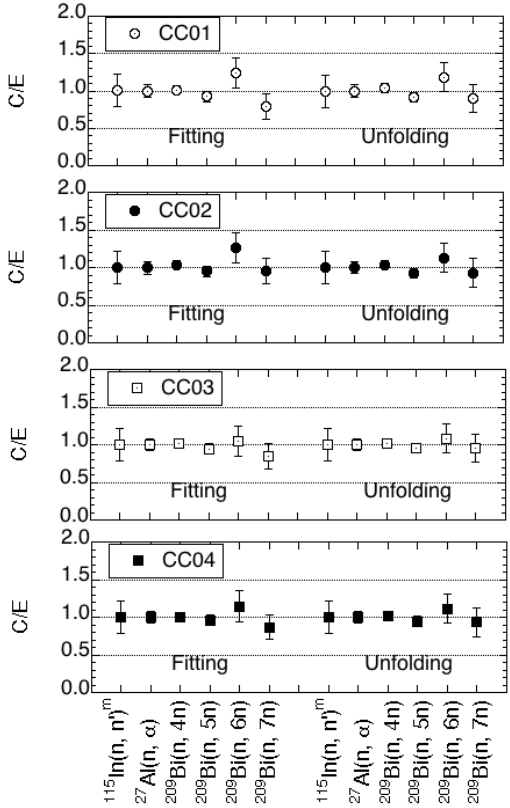

**Fig. 5.** Calculate-to-experimental reaction rate ratios. Calculated reaction rates were determined by the neutron spectra of fitting and unfolding methods shown in Fig. 4.

## 4. ANALYSES BY CALCULATION CODE

Calculation analysis was performed using a Monte Carlo code PHITS. In the calculation, we used a two-dimensional model in which the Pbar target, some of

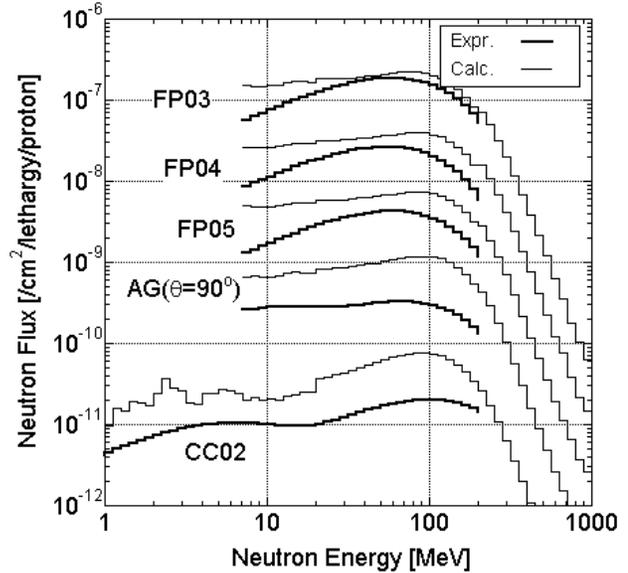

**Fig. 6.** Calculated and experimntal neutron spectra at 90° with respect to the proton beam direction. Thick lines show the measured neutron spectra, and thin lines show calculated neutron sectra. The location of the samples are given in Fig. 1.

## 5. SUMMARY

Neutron spectra in the high-energy region from 1 to 100 MeV for the shielding configuration of the Pbar target station were determined using the fitting and unfolding methods, based on the experimental reaction rate data. The neutron spectra determined by the fitting method were compared with the calculation results by the PHITS. The comparisons showed that the PHITS is suitable for shielding design of high-energy proton accelerators.

## 6. ACKNOWLEDGMENTS

This work is supported by grant aid of ministry of education (KAKENHI 19360432 and 20354764) in Japan. Fermilab is a U.S. Department of Energy Laboratory operated under Contract DE-AC02-07CH11359 by the Fermi Research Alliance, LLC.

## 7. REFERENCES

[1] H. Nakashima, Y. Sakamoto, Y. Iwamoto, N. Matsuda, Y. Kasugai, Y. Nakane, F. Masukawa, N. V. Mokhov, A. F. Leveling, D. J. Boehnlein, K. Vaziri, T. Sanami, H. Matsumura, M. Hagiwara, H. Iwase, N. Kinoshita, H. Hirayama, K. Oishi, T. Nakamura, H. Arakawa, N. Shigyo, K. Ishibashi, H. Yashima, N. Nakao and K. Niita, "Experimental studies of shielding and irradiation effects at high energy accelerator facilities," *Nucl. Technol.*, **168**, 482 (2009).

[2] H. Iwase, K. Niita and T. Nakamura, "Development of General-Purpose Particle and Heavy ion Transport Monte Carlo Code," *J. Nucl. Sci. Tecnol.*, **39**, 1142 (2002).

[3] H. Yashima, Y. Kasugai, N. Matsuda, H. Matsumura, H. Iwase, N. Kinoshita, N. Mokhov, A. Leveling, D. Boehnlein, K. Vaziri, G. Lauten, W. Schmitt, T. Nakamura, K. Oishi, H. Hirayama, K. Ishibashi, H. Nakashima and Y. Sakamoto, "Shielding Experiments at High Energy Accelerators of Fermilab (II) - Spatial distribution measurement of reaction rate behind the shield and its application for Moyer model -," *J. Nucl. Sci. Technol.*, (to be published).

[4] Y. Kasugai, H. Matsumura, H. Yashima, N. Matsuda, N. Kinoshita, H. Iwase, T. Sanami, Y. Iwamoto, M. Hagiwara, N. Sigyo, H. Arakawa, N. Mokhov, A. Leveling, D. Boehnlein, K. Vaziri, G. Lauten, W. Schmitt, V. Cupps, B. Kershisnik, S. Benesch, H. Hirayama, T. Nakamura, K. Oishi, K. Ishibashi, K. Niita, Y. Sakamoto, and H. Nakashima, "Shielding Experiments Under JASMIN Collaboration at Fermilab (I) Overview of the Research Activities," *This proceedings*.

[5] F. Maekawa, U. Möllendorf, P. Wilson, M. Wada and Y. Ikeda, "Production of a Dosimetry Cross Section Set Up to 50 MeV," *Proc of the Tenth Int. Symposium of Reactor Dosimetry*, Osaka, Japan, September, 1999.

[6] W. Mcelroy, S. Breg and G. Gigas, "Neutron-Flux Spectral Determination by Foil Activation," *Nucl. Sci. Eng.*, **27**, 533 (1967).